\documentclass[doublecol]{epl2}

\usepackage{graphicx}
\title{Dimensional crossover in Sr$_2$RuO$_4$ within slave-boson mean-field theory}
\shorttitle{Dimensional crossover in Sr$_2$RuO$_4$} 
\author{M. H. Fischer \and M. Sigrist}
\shortauthor{M. H. Fischer  \etal}

\institute{                    
Institute for Theoretical Physics, ETH Z\"urich, CH-8093 Z\"urich}


\abstract{Motivated by the anomalous temperature dependence of the c-axis resistivity of Sr$_2$RuO$_4$, the dimensional crossover from a network of perpendicular one-dimensional chains to a two-dimensional system due to a weak hybridization between the perpendicular chains is studied. 
The corresponding two-orbital Hubbard model is treated within a slave-boson mean-field theory (SBMFT) to take correlation effects into account such as the spin-charge separation on the one-dimensional chains. Using an RPA-like formulation for the Green's function of collective spinon-holon excitations the emergence of quasiparticles at low-temperatures is examined. The results are used to discuss the evolution of the spectral density and the c-axis transport within a tunneling approach. For
the latter a regime change between low- and high-temperature regime is found in qualitative accordance with experimental data. 
}

\begin{document}

\maketitle

\section{Introduction}
\begin{figure}
\onefigure{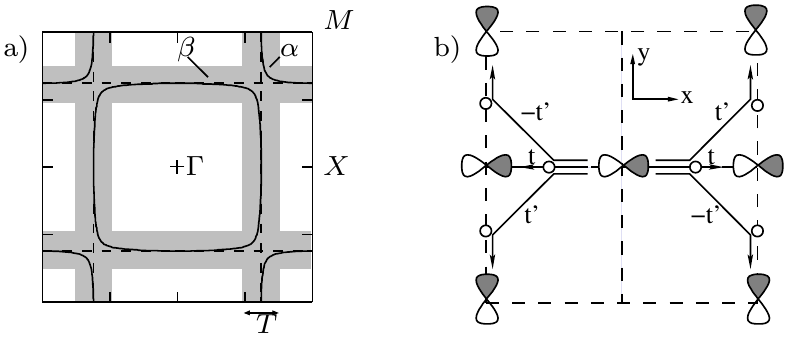}
\caption{a) The two-dimensional nature of the Fermi surfaces for the $\textrm{d}_{zx}$ and $\textrm{d}_{yz}$ orbitals is lost for temperatures above the weak inter-orbital hybridization so that the electrons behave like being in one-dimensional bands. b)~The hopping structure for the $\textrm{d}_{zx}$ and $\textrm{d}_{yz}$ electrons (here shown for a $\textrm{d}_{zx}$ placed in the middle): in lowest order via an O $2p_z$ to a nearest neighbor of the same band or as a next order term via 2 O $2p_z$ to a next-nearest neighbor of the other band.}
\label{fig:hybrid}
\end{figure}

Physical effects due to reduced dimensionality are most prominently visible for strongly correlated electrons. While under generic conditions low-energy electronic excitations retain their quasiparticle character in three and most likely two dimensions, correlated electrons in one dimension fractionalize in separate collective charge and spin excitations. All solids are three-dimensional and low-dimensionality appears through specific highly anisotropic electronic structure. 
In many cases, the effective dimensionality of such systems depends on the difference of energy scales which can lead to dimensional crossovers as systems are cooled or system parameters are changed, for example, by applying pressure. The change of effective dimensionality is often marked through the onset of ordered states which are suppressed in low-dimensional systems due to
thermal or quantum fluctuations. Such changes are well known from organic compounds, such as 
(TMTSF)$_2$X which undergoes a sequence of transitions as function of temperature, pressure and chemical composition (X). 
Another remarkable feature is the modification of transport properties which can depend on the nature of the charge carriers or on a change from coherent to incoherent transport. An example of such a system is Sr$_2$RuO$_4$ which is a quasi-two-dimensional strongly correlated Fermi liquid best known for its unconventional superconducting phase\cite{mackenzie:03}. Sr$_2$RuO$_4$ has a single-layered perovskite structure with tetragonal symmetry. Down to lowest normal state temperatures it displays a strongly anisotropic electrical resistivity: $ \rho_{c} / \rho_{ab} \sim 10^{3} $ at $ T = 2 K $. Although this system shows Fermi liquid behavior ($\rho = \rho_0 + AT^2 $) in all directions at low temperatures ($ T< 40 K $) , there is an extraordinary difference in the temperature dependence of the resistivity at higher temperature for inplane and $c$-axis transport. While the inplane resistivity is monotonically increasing, along the $c$-axis it shows a regime change from low-temperature metallic to high-temperature insulating behavior around $ T^* \approx 130 K $ \cite{lichtenberg92}. This change is not associated with any structural transition or ordering phenomenon and has been attributed to a so far unspecified correlation effect of the electrons. For the electrons to tunnel between layers, sufficient quasiparticle weight and lifetime in the planes, respectively, are required. A rapid loss of quasiparticle weight when increasing the temperature was observed in ARPES measurements~\cite{wang:04} and attributed to scattering processes. Motivated by this behavior we propose here a mechanism based on a dimensional crossover within the basal plane of Sr$_2$RuO$_4$ which affects the interlayer transport. Alternative explanations based on electron-phonon coupling~\cite{gutman2007, ho:045101} and charge fluctuations caused by the tunneling electrons~\cite{turlakov:01b} have been discussed in the literature.
The description of the electronic properties of Sr$_2$RuO$_4$ requires three bands crossing the Fermi energy with an average filling of $4/3$. Their shape is well accounted for by considering the $ \pi $-hybridization of the O-$2p$-orbitals with $4d$-$t_{2\textrm{g}}$-orbitals of Ru in each ruthenium oxide layer\cite{Oguchi:95, Singh:95}. The $d_{xy}$-orbital lies in the RuO$_2$ plane and forms the genuinely two-dimensional $\gamma$-band. The other two orbitals $ d_{yz} $ and $ d_{zx} $ forming the so-called $ \alpha $- and $ \beta $-band do not hybridize with the $ \gamma $-band as they have different parity with respect to reflection at the basal plane. In contrast to the $ \gamma $-band which has practically no interlayer coupling, these other two orbitals are also responsible for the weak interlayer hybridization~\cite{bergemann:2000}. Thus, in order to understand the $c$-axis transport we will focus our attention to the $ d_{yz} $- and $ d_{zx} $-orbitals which have a peculiar hybridization topology within the basal plane. Both orbitals form through nearest-neighbor intra-orbital $ \pi $-hybridization an essentially one-dimensional band dispersion along the $ y $-direction ($x$-direction) for the $ d_{yz} $-orbital ($d_{zx} $-orbital). The inter-orbital coupling occurs through considerably weaker next-nearest-neighbor hybridization between $ d_{yz} $ and $ d_{zx} $ and leads to two-dimensional bands which nevertheless keep strong one-dimensional features (see fig. ~\ref{fig:hybrid}a)) with pronounced nesting properties reflected in the enhanced incommensurate magnetic spin fluctuations~\cite{Mazin:99}.  We show here that the structure of these orbitals could lead to a crossover behavior between the one- and two-dimensional regime at the energy scale associated with the next-nearest-neighbor inter-orbital coupling similar to other systems discussed in literature~\cite{Giamarchi:2004}. For this purpose we develop a slave-boson scheme on the model of the $ d_{yz} $-$d_{zx}$-orbitals in order to describe the fractionalization of the quasiparticles in the one-dimensional system. We will show that this method allows us to qualitatively account for the emergence of quasiparticle states in the hybridized bands in a rather simple manner. This method will also be used to give a qualitative discussion of the regime change in the interlayer transport which we consider in the tunneling limit. 
\section{Formalism}
We start with the two independent, one-dimensional bands (henceforth labeled $\nu=1,2$ corresponding to the $x$- and $y$-directions of the $d_{zx} $- and $ d_{yz} $-orbitals, respectively) described by the Hubbard Hamiltonian
\begin{equation}
    \mathcal{H} = -t \sum_{\phantom{\langle}\!\!\nu, \sigma}\sum_{\langle i,j \rangle_{\nu}} c^{\dag}_{\nu i \sigma} c^{\phantom{dag}}_{\nu j \sigma} + U\sum_{\phantom{\langle}\!\!\nu, i}n^{\phantom{\dag}}_{\nu i \uparrow} n^{\phantom{\dag}}_{\nu i \downarrow}
    \label{ham-1}
\end{equation}
with $t$ the hopping integral, $U$ the on-site Coulomb interaction and $\langle i, j\rangle_{\nu}$ denoting nearest neighbors in the corresponding direction. The operator $c^{\dag}_{\nu i \sigma}$ creates an electron in the orbital $\nu$ at site $i$ with spin $\sigma$ and $n^{\phantom{\dag}}_{\nu i \sigma} =c^{\dag}_{\nu i \sigma}c^{\phantom{\dag}}_{\nu i \sigma}$. Since such correlated electron systems are difficult to handle we seek an approximation by assuming that Coulomb repulsion is sufficiently strong as to suppress essentially double occupancy. In this case we may use a slave-boson technique which had been developed in the context of the $t$-$J$-model.  By writing the electron operators $c_{\nu i\sigma}$ in terms of new fermionic operators representing the spin degrees of freedom (spinons) and bosonic operators for the charge degrees of freedom (holons), $c^{\phantom{\dag}}_{\nu i\sigma}=f^{\phantom{\dag}}_{\nu i\sigma}b^{\dag}_{\nu i}$ and stipulating that each site is either occupied by a spinon or a holon, double-occupancy is strictly inhibited~\cite{RUCKENSTEIN:1987}. From this replacement the so-called slave-boson Hamiltonian results,
\begin{eqnarray}
	\mathcal{H} &=& -t\sum_{\phantom{\langle}\!\!\nu, s}\sum_{\langle i,j \rangle_{\nu}} f^{\dag}_{\nu i s}b^{\phantom{\dag}}_{\nu i} b^{\dag}_{\nu j} f^{\phantom{\dag}}_{\nu j s}\nonumber \\
	&& + \sum_{\nu i}\lambda^{\phantom{\dag}}_{\nu i}\left(b^{\dag}_{\nu i}b^{\phantom{\dag}}_{\nu i} + \sum_s f^{\dag}_{\nu i s}f^{\phantom{\dag}}_{\nu i s} - 1\right)\nonumber\\
	&& - \mu \sum_{\nu, i, s} f^{\dag}_{\nu i s}f^{\phantom{\dag}}_{\nu i s}
\end{eqnarray}
with the Lagrange multipliers $\lambda_{\nu i} $ enforcing the constraints of one spinon or holon per site. To analyze this Hamiltonian we resort to the following standard approximations. We decouple the first term by introducing mean-fields for the coherent motion of holons and spinons as
\begin{equation}\label{eq:meanfields}
	\chi_i^{\nu b} = \langle b^{\dag}_{\nu i} b^{\phantom{\dag}}_{\nu j}\rangle\quad \textrm{and} \quad \chi_i^{\nu f} = \sum_{\sigma}\langle f^{\dag}_{\nu i\sigma} f^{\phantom{\dag}}_{\nu j\sigma}\rangle,
\end{equation}
respectively, with $i$, $j$ being nearest-neighbor indices. The treatment of the Hamiltonian is then further simplified by taking the mean-fields as being independent of band and site and replacing the local constraints by a global one.

With this approximation we arrive at  a slave-boson mean-field Hamiltonian representing two independent species of particles, spinons and holons, in two independent one-dimensional bands
\begin{equation}\label{eq:h0}
    \mathcal{H}_{0} = \sum_{\nu, \vec{k}, \sigma}\varepsilon^{\phantom{\dag}}_{\nu\vec{k}} f^{\dag}_{\nu\vec{k} \sigma}f^{\phantom{\dag}}_{\nu\vec{k} \sigma} + \sum_{\nu, \vec{k}}\omega^{\phantom{\dag}}_{\nu\vec{k}}b^{\dag}_{\nu\vec{k}}b^{\phantom{\dag}}_{\nu\vec{k}}
\end{equation}
with $\varepsilon_{\nu\vec{k}} = - 2t \chi^{b}\cos(k_{\nu}) + \lambda - \mu$ the spinon energy and $\omega_{\nu\vec{k}} = - 2t  \chi^{f} \cos(k_{\nu}) + \lambda$ the holon energy. Fixing $x$, the average number of holons per site (thus also the average number of spinons) and using eq.~(\ref{eq:meanfields}), we can then self-consistently determine the values of $\lambda$, $\mu$, $\chi^f$ and $\chi^b$. In this way we incorporate effectively the spin-charge separation of the one-dimensional correlated electron system realized in
the nearest-neighbor hopping Hamiltonian (\ref{ham-1}). We now add the  next-nearest-neighbor interband hopping via two O $2p$-orbitals as displayed in fig.~\ref{fig:hybrid}b). This weaker hopping term connects different one-dimensional electron systems and may be written in the slave-boson representation as
\begin{equation}
    \mathcal{H}' = \sum_{\phantom{\vec{k}}\!\!\!\!\sigma}\sum_{\vec{k}, \vec{k}',\vec{q}}\left(g^{\phantom{\dag}}_{\phantom{\vec{k}}\!\!\!\!\vec{q}} f^{\dag}_{1 \vec{k} + \vec{q} \sigma}b^{\phantom{\dag}}_{1 \vec{k}} b^{\dag}_{2\vec{k}'} f^{\phantom{\dag}}_{2 \vec{k}' + \vec{q} \sigma}+h.c.\right)
\end{equation}
with the structure factor $g_{\vec{q}} = -4t'\sin(q_x)\sin(q_y)$ and $t'$ the interband hopping integral. We now use the fact that between these systems only electrons can be transfered, invoking that a spinon and holon have to correlate to yield hopping. Thus we may interpret $  \mathcal{H}' $ as an effective interaction term introducing attractive coupling between the two subspecies with the tendency to recombine them to an electronic quasiparticle~\cite{Ng:2005}. 

We now investigate the electronic spectrum using the retarded single-electron Green's function defined as
\begin{equation}
    G^{\nu\nu'}\!(\vec{q}, t) = -\Theta(t)\langle\{c^{\phantom{\dag}}_{\nu\vec{q}\sigma}(t), c^{\dag}_{\nu'\vec{q}\sigma}(0)\}\rangle.
\end{equation}
where the electron operators can be replaced by the spinon and holon operators,
\begin{equation}\label{eq:electron}
    c^{\dag}_{\phantom{\vec{k}}\!\!\!\!\nu \vec{q} \sigma} = \frac{1}{N}\sum_{\vec{k}}f^{\dag}_{\nu \vec{k} + \vec{q} \sigma} b^{\phantom{\dag}}_{\nu\vec{k}}
\end{equation}
and $\Theta(t)$ is the Heaviside step-function. In eq.~(\ref{eq:electron}), $N$ is the number of lattice sites in one direction.\\
Following standard methods we first calculate an auxiliary Green's function
\begin{equation}
    g^{\nu\nu'}_{\vec{k}}\!(\vec{q}, t) = -\Theta(t)\langle \{b^{\dag}_{\nu\vec{k}}(t)f^{\phantom{\dag}}_{\nu \vec{k} + \vec{q} \sigma}(t),c^{\dag}_{\nu' \vec{q} \sigma\phantom{\vec{k}}\!\!}(0)\}\rangle
\end{equation}
which can be evaluated in energy space using the equation of motion,
\begin{eqnarray}
    E g^{\nu\nu'}_{\vec{k}}\!(\vec{q}, E) &=& \frac{1}{N}\sum_{\vec{p}}\left\langle\left\{b^{\dag}_{\nu\vec{k}}f^{\phantom{\dag}}_{\nu\vec{k} + \vec{q} \sigma},f^{\dag}_{\nu' \vec{p} + \vec{q} \sigma\phantom{\vec{k}}\!\!\!\!} b^{\phantom{\dag}}_{\nu'\vec{p}\phantom{\vec{k}}\!\!\!\!}\right\}\right\rangle\nonumber\\
    && -\left\langle\left[\mathcal{H},b^{\dag}_{\nu \vec{k}}f^{\phantom{\dag}}_{\nu \vec{k} + \vec{q} \sigma}\right],c^{\dag}_{\nu' \vec{q} \sigma\phantom{\vec{k}}\!\!\!\!}\right\rangle.
\end{eqnarray}
This equation of motion involves commutators of the form
\begin{equation}
    \left[\mathcal{H}_0, b^{\dag}_{\nu\vec{k}}f^{\phantom{\dag}}_{\nu \vec{k} + \vec{q} \sigma}\right] =- E^{\nu\nu}_{\vec{k} + \vec{q}, \vec{k}}b^{\dag}_{\nu\vec{k}}f^{\phantom{\dag}}_{\nu \vec{k} + \vec{q} \sigma}
\end{equation}
with $E^{\nu\nu}_{\vec{k} + \vec{q}, \vec{k}} = \varepsilon^{\phantom{\nu}}_{\nu\vec{k}+\vec{q}} - \omega^{\phantom{\nu}}_{\nu\vec{k}}$ and
\begin{eqnarray}\label{eq:comm2}
    \left[\mathcal{H}', b^{\dag}_{\nu\vec{k}}f^{\phantom{\dag}}_{\nu \vec{k} + \vec{q} \sigma}\right] \!\!\!\!&=&\!\!\!\!\! \sum_{\vec{k}',\vec{q}', \sigma'}g^{\phantom{\dag}}_{\vec{q}'\phantom{\vec{k}}\!\!\!\!}f^{\dag}_{\nu \vec{k} + \vec{q}' \sigma'}f^{\phantom{\dag}}_{\nu \vec{k} + \vec{q} \sigma} b^{\dag}_{\bar{\nu} \vec{k}'}f^{\phantom{\dag}}_{\bar{\nu} \vec{k}' + \vec{q}'\sigma'}\nonumber\\
    &&\!\!\!\!\!+ \sum_{\vec{k}',\vec{q}'}g^{\phantom{\dag}}_{\vec{q}'\phantom{\vec{k}}\!\!\!\!}b^{\dag}_{\nu \vec{k}}b^{\phantom{\dag}}_{\nu \vec{k}+\vec{q} - \vec{q}'} b^{\dag}_{\bar{\nu} \vec{k}'}f^{\phantom{\dag}}_{\bar{\nu} \vec{k}' + \vec{q}'\sigma}.
\end{eqnarray}
The higher order Green's functions coming from (\ref{eq:comm2}) are treated within RPA to yield
\begin{equation}
    \hat{G}(\vec{q}, E) = \hat{G}_0(\vec{q}, E) + \hat{G}_0(\vec{q}, E)\hat{g}(\vec{q}) \hat{G}(\vec{q}, E)
\end{equation}
where
\begin{equation}
    \hat{G}_0(\vec{q}, E) = \left(\begin{array}{cc} G_0^1(\vec{q}, E) & 0 \\ 0 & G_0^2(\vec{q}, E)\end{array}\right),
\end{equation}
\begin{equation}\label{eq:sbmft_bare}
    G_0^{\nu}(\vec{q}, E) = \frac{1}{N^2}\sum_{\vec{k}} \frac{n_{F}^{(\nu)}(\vec{k} + \vec{q}) + n_{B}^{(\nu)}(\vec{k})}{E - E^{\mu\mu}_{\vec{k} + \vec{q}, \vec{k}}}
\end{equation}
is the bare Green's function resulting from the Hamiltonian (\ref{eq:h0}) without inter-band hopping and
\begin{equation}
    \hat{g}(\vec{q}) = \left(\begin{array}{cc} 0 & g_{\vec{q}} \\ g_{\vec{q}} & 0 \end{array}\right).
\end{equation}
Eventually, we find the RPA form of the single-electron Green's function in SBMFT
\begin{equation}\label{eq:sbmft_greens}
    G^{\nu\nu}(\vec{q}, E) = \frac{G_0^{\nu}(\vec{q}, E)}{1 - \left(g_{\vec{q}}\right)^2G_0^1(\vec{q}, E)G_0^2(\vec{q},E)}.
\end{equation}

\section{Results}
\begin{figure}
\onefigure{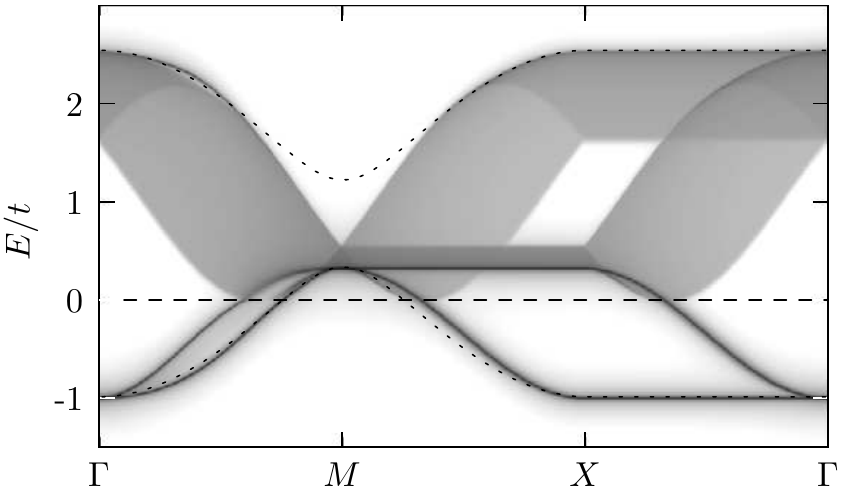}
\caption{Density log-plot of the spectral density function along the main symmetry axis for $T=0.005t$ showing the dispersion of the quasiparticles ('electrons'). The dotted lines mark the boundaries of the spinon-holon continuum. The actual quasiparticle band is thus strongly renormalized. The dashed, white horizontal line denotes the Fermi energy.}
\label{fig:band}
\end{figure}
\begin{figure}
\onefigure{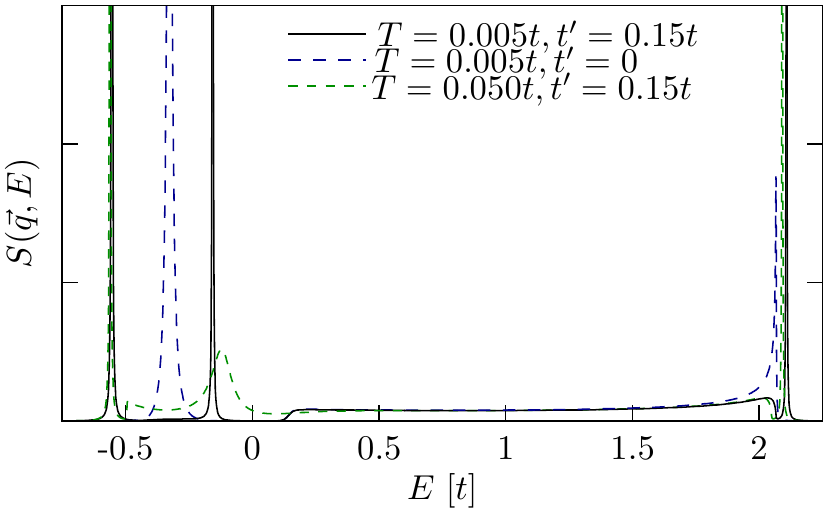}
\caption{Comparison of the spectral density at $\vec{q}=(\pi/2, \pi/2)$ as a function of energy for the cases of low and high temperatures with interband hopping and low temperature without interband hopping.}
\label{fig:scan}
\end{figure}
The total spectral density in momentum-space associated with the Green's function (\ref{eq:sbmft_greens}),
\begin{equation}\label{eq:specfct}
    S(\vec{q},E) = -\frac{1}{\pi} {\rm tr} \{\textrm{Im} (G^{\nu\nu}(\vec{q},E)]\},
\end{equation}
is evaluated numerically where the hole concentration is $x=0.33$ and the inter-band hopping is $t'/t = 0.15$ as parameters fitting to the de Haas-van Alphen Fermi surface~\cite{Bergemann:2003}.
For $\vec{q}$ along the $x$ or the $y$ axis in momentum space, $g_{\vec{q}}$ vanishes and the denominator of (\ref{eq:sbmft_greens}) is equal to 1 without any modifications to the original one-dimensional situation.  
In contrast, for $\vec{q}$ along the diagonal, $|g_{\vec{q}}|$ is maximal and the renormalization of the Green's function is strongest. Indeed the recombination of spinons and holons into quasiparticles is here most pronounced, visible in the band hybridization and the evolution of quasiparticle weight as we will see below. 

First, we examine the spectrum along the main symmetry axis in the Brillouin zone (see fig.~\ref{fig:band}). At high temperatures, this spectrum is represented by the two-particle continuum of a spinon and a holon, whose boundary is marked by the dotted lines in fig.~\ref{fig:band}. At low temperatures the spectral weight shifts into two quasiparticle bands with a sharp momentum energy relation. These bands have the shape which is expected from a simple tight-binding model taking nearest- and next-nearest-neighbor hybridization into account. Note that we ignore the lifetime effect due to the scattering among the resulting quasiparticles. Thus, these quasiparticle states remain well defined even away from the Fermi energy.
The shift of spectral weight to the quasiparticle bands is further illustrated in fig.~\ref{fig:scan} which shows a cut of the spectral density at $\vec{q} = (\pi/2, \pi/2)$ for different temperatures and interband couplings. We see in this plot that the peak around $E=-0.3t$ without an interband coupling splits into two well-defined peaks of which the one at lower energies stays even for higher temperatures while the one at higher energies gets washed out.

\begin{figure}
\onefigure{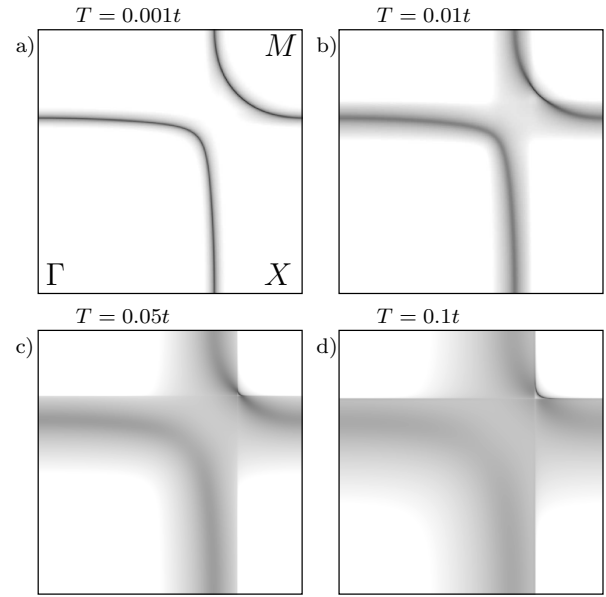}
\caption{Density log-plots of the spectral density in momentum space at the Fermi energy for different temperatures. For $T=0.001t$, we see a Fermi surface as expected for a hybridized, two-dimensional system. When increasing the temperature the peaks of the spectral function get smeared and eventually, for $T=0.1t$ we find a picture similar to that of a system of two decoupled one-dimensional bands.}
\label{fig:fs}
\end{figure}

\begin{figure}
\onefigure{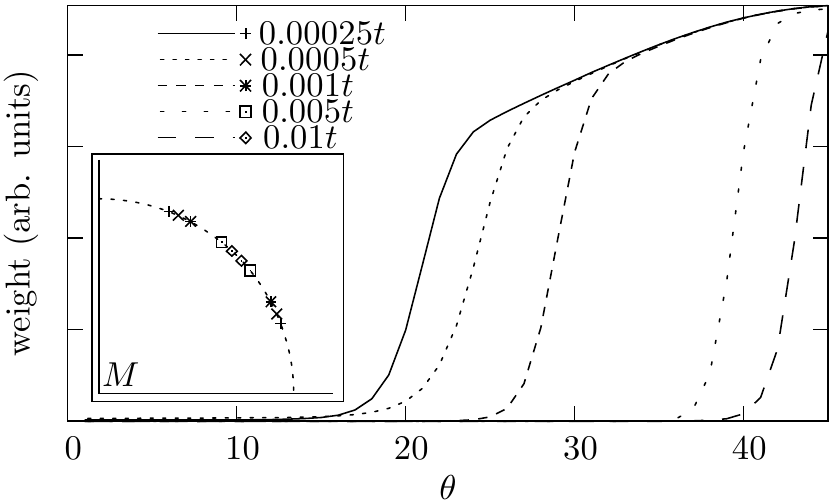}
\caption{Quasi-particle weights as calculated from the spectral density function in arbitrary units around the $M$-point as a function of the angle to the $x$-axis for different temperatures. When decreasing the temperature a growing arc of quasiparticle weight appears around the diagonals. To emphasize this, the inset shows the endpoints of the arcs for different temperatures.}
\label{fig:qp}
\end{figure}

Second, we further analyze the $q$-dependence as well as the temperature dependence of the quasiparticle spectral weight. Figure~\ref{fig:fs} shows density log-plots of the spectral function at the Fermi energy for different temperatures. At lowest temperatures ($ T = 0.001 t $) the two-dimensional Fermi surface is well defined by quasiparticles (fig.~\ref{fig:fs}a)). With increasing temperature, the quasiparticle weight decreases and the Fermi surface becomes gradually faint. The retreat of the quasiparticle weight is clearly away from the [100] and [010] directions towards the [110] direction (figs.~\ref{fig:fs}.b-d)). A similar behavior of the quasiparticle weight was observed in ARPES experiments~\cite{wang:04}. However, only the $\gamma$-band was studied in the [110] direction and thus, a complete comparison of the results is beyond the scope of our approach.\\
In our calculation, quasiparticles occur first along the diagonal momentum direction and extend slowly over a wider range of the emerging Fermi surface to form in an intermediate stage pieces of arc-like Fermi surfaces. 
To illustrate the evolution of quasiparticles at the Fermi surface,  in fig.~\ref{fig:qp} we show the angle dependence of the quasiparticle weight of the pocket around the $M$-point for different temperatures. This clearly demonstrates the gradual build-up of quasi particle weight as the inset of fig.~\ref{fig:qp} shows. Within our calculation the Fermi arcs appear only clearly for the Fermi surface around the $M$-point ($\alpha$-Fermi surface). While for the Fermi surface centered around the $ \Gamma $-point the rise of quasiparticle weight does not involve such a well defined arc feature.\\
Note that unlike in other slave-boson discussions, here the holon condensation is not involved in the formation of coherent quasiparticles. The formation of the quasiparticles in the spectrum is caused by the inter-orbital hybridization which is taken into account in the RPA-like form of the Green's function~(\ref{eq:sbmft_greens}). The picture obtained in our slave-boson mean field approach should agree on a qualitative level with other descriptions of the dimensional crossover starting out from a Luttinger-liquid in one dimension. The slave-boson mean-field theory carries the advantage of formal simplicity.

\section{$c$-axis transport}
\begin{figure}
\onefigure{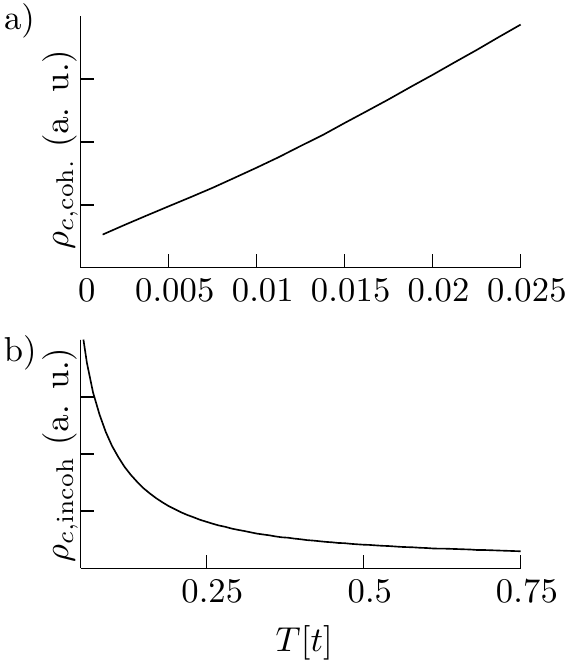}
\caption{The resistivity in the low- and high-temperature limit as calculated from the eqs.~(\ref{eq:coh}) and (\ref{eq:incoh}).a) As expected, the decreasing quasiparticle weight in the low-temperature limit results in an increase in the resistivity as the temperature is increased. b) In the high-temperature limit, the tunneling possibility of the spinons dominates the picture resulting in a decreasing resistivity.}
\label{fig:res}
\end{figure}

For the transport along the $c$-axis we assume that the interlayer tunneling approach is justified due to the extremely large anisotropy of resistivity. We introduce the tunneling Hamiltonian 
\begin{equation}
    \mathcal{H}_{\rm{T}} = \frac{1}{N^2}\sum_{\nu\nu'\sigma\phantom{\vec{k}}\!\!\!\!}\sum_{\vec{k}, \vec{k}', \vec{p}, \vec{p}'}(T_{\vec{k} \vec{p}}^{\nu\nu'\phantom{\dag}\!\!\!\!}f^{\dag}_{\nu \vec{k}' + \vec{k}\sigma}b^{\phantom{\dag}}_{\nu \vec{k}'}b^{\dag}_{\nu' \vec{p}'\phantom{\vec{k}}\!\!\!\!}f^{\phantom{\dag}}_{\nu' \vec{p}' + \vec{p} \sigma\phantom{\vec{k}}\!\!\!\!} + h.c.)
\end{equation}
where operators with momentum $\vec{k}$ belong to one plane and with $\vec{p}$ to the other.
The matrix elements are 
\begin{equation}
	T_{\vec{k} \vec{p}}^{\nu\nu'} = 4 t'' \delta^{\phantom{\nu}}_{\vec{k}\vec{p}} \left\{ \delta^{\phantom{\nu}}_{\nu \nu'} \cos \frac{k_x+k_y}{2} + \sin \frac{k_x}{2} \sin \frac{k_y}{2} \right\}
\end{equation}
for hopping from one band to the same and from one band to the other, respectively. This form results from the analysis of the possible matrix elements between different layers taking symmetry aspects into account. The total tunneling current can be expressed as the sum of a coherent and an incoherent tunneling current
\begin{equation}
	I_{\rm{tot}}=I_{\rm{coh}} + I_{\rm{inc}}. 
\end{equation}
Depending on the temperature range we expect one or the other part to dominate the transport. The maximum in $c$-axis resistivity in Sr$_2$RuO$_4$ is around $ T =130 K $, for $t\approx0.3$eV~\cite{Bergemann:2003} this corresponds to a temperature of $T\approx0.05t$. We therefore have to study a low-temperature region with $T\ll0.05t$ and a 'high'-temperature region with $T\gg0.05t$.\\
In the low temperature limit, we saw that the quasiparticle weight of the coherent spinon-holon pairs is growing with decreasing temperature and we expect the main contribution to the tunneling current~\cite{schrieffer:63} to come from the coherent component,
\begin{equation}\label{eq:coh}
    I = \frac{2e^2}{\hbar}V\sum_{\nu\nu'\phantom{\vec{k}}\!\!\!}\sum_{\vec{k} \vec{p}}|T_{\vec{k}\vec{p}}^{\nu\nu'}|^2 S_{\rm{U}\phantom{\vec{k}}\!\!\!}^{\nu}\!(\vec{k}, 0)S_{\rm{L}\phantom{\vec{k}}\!\!\!}^{\nu'}\!(\vec{p}, 0).
\end{equation}
Here, $S_{\rm{L}(\rm{U})}^{\nu}$ denotes the spectral density of the band $\nu$ in the lower (upper) layer. Using the spectral density as calculated from eq.~(\ref{eq:specfct}) this indeed yields an increasing resistivity as is shown in fig.~\ref{fig:res}a), because of the diminishing of quasiparticle weight upon growing temperature. Note that in this discussion we ignore any other source of temperature dependence.  \\
In the high-temperature limit, we see from figs.~\ref{fig:fs} and \ref{fig:qp} that coherent quasiparticles  that could tunnel between layers have disappeared. Therefore, the quasiparticle picture based on paired spinons and holons as used for the discussion of the tunneling current above is inappropriate. We have to take a different approach whereby spin and charge degrees of freedom are independent. In the tunneling process of an electron, a spinon and a holon (charge carrier) have to be transfered simultaneously giving rise to an incoherent transport~\cite{zou:88}. 
We start with the Fermi Golden Rule to estimate the transfer probability of holons,
\begin{equation}
    \Gamma^{b}_{i\rightarrow j} = \frac{2\pi}{\hbar}\sum_{\phi'}|\langle\phi_0|\mathcal{H}_T|\phi'\rangle|^2\delta(E_0 - E' + eV)
\end{equation}
with $\phi_0$ the many-body ground-state, $\phi'$ any excited state and $E_0$ and $E'$ the corresponding energies. Decoupling the holons and the spinons, we find for the total rate of holon tunneling
\begin{eqnarray}
	\Gamma^{b}_{i\rightarrow j}\!\! &=&\!\! 8 \frac{2\pi}{\hbar}|T_{\bot}|^2\sum_{k, k', q}n_F^{(j)}(q)[1 - n_F^{(i)}(q)]\times\nonumber\\
	&&\times n_B^{(i)}(k')[1+n_B^{(j)}(k)]\delta(\omega_{k} - \omega_{k'}+eV).\label{eq:gammak}
\end{eqnarray}
which requires that a spinon moves in the direction opposite to the holon. 
Here, $n_{F(B)}^{(i)}(k)$ denotes the Fermi (Bose) distribution with momentum $k$ in the layer $i$, $|T_{\bot}|^2$ is the (momentum independent) tunneling matrix element. The factor of $8$ in the above formula is composed of a factor of 2 for spins and a factor of 4 because of the different bands involved. Since the distributions are the same on both layers, we can write the net tunneling rate as
\begin{eqnarray}
    &&\Gamma^{b}_{\rm{net}} = \frac{2\pi}{\hbar}8|T_{\bot}|^2 \Big(\int d\epsilon D_F(\epsilon) n_F(\epsilon)[1-n_F(\epsilon)]\Big)\times\nonumber\\ &&\times\!\!\int\!\! d\omega D_B(\omega)D_B(\omega+ eV)[n_B(\omega + eV)-n_B(\omega)]\label{eq:netprob}
\end{eqnarray}
where $D_{F(B)}$ is the density of states of the spinons (holons). The integral over the Fermi distribution functions yields a factor of $D_F T$ with $D_F$ the density of states of the spinons taken to be constant at the spinon Fermi level. Taking the density of states of the holons, $D_B$,  to be constant too, the integral over the Bose distributions can be written as
\begin{equation}
	(D^{\phantom{2}}_B)^2 \int d\omega \{ n^{\phantom{2}}_B(\omega + eV)-n^{\phantom{2}}_B(\omega)\} \approx D^{\phantom{2}}_Bn_B^2 \kappa^{\phantom{2}}_BeV,
\end{equation}
with the compressibility of the hard-core holons, $\kappa_B$, the chemical potential, $\mu$ and $n_B$ the holon density.\\
Based on eq.~(\ref{eq:netprob}) the resulting current can now be estimated as
\begin{equation}
    I = 8\frac{e^2}{\hbar}(2 \pi)|T^{\phantom{2}}_{\bot}|^2 D^{\phantom{2}}_F D^{\phantom{2}}_B T n_B^2\kappa^{\phantom{2}}_B(T) V
\end{equation}
leading to the conductivity
\begin{equation}\label{eq:incoh}
    \sigma = 8 \frac{e^2}{\hbar}(2 \pi)|T^{\phantom{2}}_{\bot}|^2 D^{\phantom{2}}_F D^{\phantom{2}}_B T n_B^2\kappa^{\phantom{2}}_B(T) \frac{c}{ab}.
\end{equation}
The compressibility of the holons, $ \kappa_B(T) $, corresponds to that of spinless Fermions and is constant for temperatures much smaller than the characteristic energy of holons which is of the order $ t \chi^f $. Thus, the temperature dependence of $ \sigma $ in the range of interest is dominated by $ T $ due to the spinon phase space contribution which is increasing with growing temperature. Therefore we find a decreasing resistivity in the temperature range above $ T = 0.05 t $. \\
In addition to the c-axis resistivity, the analysis of the magnetic susceptibility displays signatures of a dimensional crossover, as magnetic excitations due to nesting features, pronounced in one dimension, are weakened when lowering the temperature. A behavior of such kind was experimentally reported in \cite{braden:02}.

\section{Conclusion}
In our discussion we have interpreted the regime change of the $c$-axis transport in Sr$_2$RuO$_4$ in terms of a dimensional crossover (from one to two dimensions) for the electronic states in the basal plane. We have shown that the slave-boson approximation can give a qualitative understanding and  provides interesting insight into the change of transport properties. In particular, our results lead to a picture where with decreasing temperature a Fermi surface appears through gradually extending arcs of finite quasiparticle weight. This kind of Fermi surface evolution could be accessible to more specific ARPES experiments.
In view of the progressive emergence of coherent quasiparticles the difference between the metallic and insulating behavior can be understood
as a change between dominant coherent to dominant incoherent tunneling as temperature is raised.

\acknowledgments
We would like to thank G. Baskaran, A. Furusaki, F. Hassler, M. Indergand, L. Pollet and T.M. Rice for fruitful discussions. This study has been financially supported by the Swiss Nationalfonds and the NCCR MaNEP.

\bibliographystyle{eplbib.bst}
\bibliography{comments.bib,ref.bib}

\end{document}